\documentclass[11pt, preprint]{aastex}

\shorttitle{Widespread glycolaldehyde and ethylene glycol around Sgr B2}
 \shortauthors{Li et al.}

\begin{document}

\title{Widespread Presence of Glycolaldehyde and Ethylene Glycol Around Sagittarius B2}

\author{Juan Li\altaffilmark{1, 2}, Zhiqiang Shen\altaffilmark{1, 2}, Junzhi Wang\altaffilmark{1, 2}, Xi Chen\altaffilmark{1, 2}, Di Li\altaffilmark{2, 3}, Yajun Wu\altaffilmark{1, 2},
Jian Dong\altaffilmark{1, 2}, Rongbing Zhao\altaffilmark{1, 2}, Wei Gou\altaffilmark{1, 2}, Jinqing Wang\altaffilmark{1, 2}, Shanghuo Li\altaffilmark{1, 4}, Bingru Wang\altaffilmark{2, 3, 4}, Xingwu Zheng\altaffilmark{2, 5,6}
}

\altaffiltext{1}{Department of Radio Science and Technology, Shanghai Astronomical Observatory, 80 Nandan RD, Shanghai 200030, China; lijuan@shao.ac.cn}

\altaffiltext{2}{Key Laboratory of Radio Astronomy, Chinese Academy of Sciences, China}

\altaffiltext{3}{National Astronomical Observatories, Chinese Academy of Sciences, A20 Datun Road, Chaoyang District, Beijing 100012, China}

\altaffiltext{4}{University of Chinese Academy of Sciences, 19A Yuquanlu, Beijing, 100049, China}
 
\altaffiltext{5}{School of Astronomy and Space Science, Nanjing University, Nanjing 210093, China}

\altaffiltext{6}{Key Laboratory of Modern Astronomy and Astrophysics (Nanjing University), Ministry of Education, Nanjing 210093, China}

\begin{abstract}


We report the detection of widespread CH$_2$OHCHO and HOCH$_2$CH$_2$OH emission in Galactic center giant molecular cloud Sagittarius B2 using the Shanghai Tianma 65m Radio Telescope. Our observations show for the first time that the spatial distribution of these two important prebiotic molecules extends over 15 arc-minutes, corresponding to a linear size of approximately 36 pc. These two molecules are not just distributed in or near the hot cores. The abundance of these two molecules seems to decrease from the cold outer region to the central region associated with star-formation activity. Results present here suggest that these two molecules are likely to form through a low temperature process. Recent theoretical and experimental studies demonstrated that prebiotic molecules can be efficiently formed in icy grain mantles through several pathways. However, these complex ice features cannot be directly observed, and most constraints on the ice compositions come from millimeter observations of desorbed ice chemistry products. These results, combined with laboratory studies, strongly support the existence of abundant prebiotic molecules in ices. 

\end{abstract}


\section{Introduction}
\qquad

Prebiotic molecules are directly related to the origin of life on Earth (Herbst \& van Dishoeck 2009), thus understanding the property and formation mechanism of prebiotic molecules is key for study of astrobiology. It has been found that complex organic molecules (COMs) such as glycolaldehyde and ethylene glycol are very abundant in the Galactic center giant molecular cloud Sagittarius B2 (Sgr B2) (Hollis et al. 2000, 2002; Menten 2011). Sgr B2 contains two main sites of star formation, Sgr B2(N) and Sgr B2(M), which have been the best hunting ground for prebiotic molecules in the interstellar medium (ISM) since the early 1970s (Belloche et al. 2013). Most of the COMs, like the simplest sugar-related molecule glycolaldehyde, and the simplest polyol ethylene glycol, were first detected in Sgr B2(N) (Hollis et al. 2000, 2002, 2004). 

Glycolaldehyde, CH$_2$OHCHO, which is a sugar-related molecule, can react with propenal to form ribose - a central constituent of RNA (Sharma et al. 2016). Besides Sgr B2(N), glycolaldehyde has been detected toward only a few sources in the Galactic disk, such as solar-type protostars IRAS 16293-2422 (J${\o}$rgensen et al. 2012, 2016), NGC 1333 IRAS2A (Coutens et al. 2015), NGC1333 IRAS4A (Taquet et al. 2015), protostellar shock region L1157-B1 (Lefloch et al. 2017), massive hot cores G31.41+0.31 (Beltr$\acute{a}$n et al. 2009), intermediate-mass protostar NGC 7129 FIRS 2 (Fuente et al. 2014), and comet C/2014 Q2 (Lovejoy) (Biver et al. 2015). Ethylene glycol, HOCH$_2$CH$_2$OH, is a dialcohol, a molecule chemically related to standard alcohol (ethanol, CH$_3$CH$_2$OH). It is commonly known as an antifreeze coolant for car engines. Laboratory experiments showed that ethylene glycol is formed by hydrogenation of glycolaldehyde (Fedoseev et al. 2015). Ethylene glycol has been detected in most of the regions where glycolaldehyde was detected (J${\o}$rgensen et al. 2016; Coutens et al. 2015; Fuente et al. 2014; Biver et al. 2015; Rivilla et al. 2017) as well as in four more massive star-forming regions, including Orion KL, W51/e2, G34.3+0.2, IRAS 20126+4104 (Brouillet et al. 2015; Lykke et al. 2015; Palau et al. 2017), and two comets (Biver et al. 2015). In spiral-arm clouds, glycolaldehyde and ethylene glycol emissions are always compact and come from dense regions close to the protostars (Beltran et al. 2009; Brouillet et al. 2015), or from the cavities of outflows (Palau et al. 2017), with the angular sizes of the emission regions less than 1.2 arc-second (J${\o}$rgensen et al. 2016: Coutens et al. 2015; Palau et al. 2017; Beltran et al. 2009). 

By observing the strongest known glycolaldehyde transition with the Berkeley-Illinois-Maryland Association (BIMA) array, Hollis et al.(2001) produced a distribution map of glycolaldehyde that revealed a weak concentration of emission confined to the Large Molecule Heimat (LMH) core in Sgr B2(N) (Kuan et al. 1996), indicating that the bulk of the glycolaldehyde emission comes from the less dense cloud extremities in the vicinity of the LMH core and has a spatial scale $\geq$ 60\arcsec. However, the exact extent of glycolaldehyde and ethylene glycol still remains unclear. In this paper, we report the detection of widespread glycolaldehyde and ethylene glycol emission in Sgr B2 with the Shanghai Tianma 65m Radio Telescope (TMRT). 

\section{OBSERVATIONS AND DATA ANALYSIS }
\label{observation}

\subsection{Observations and data reduction}

Mapping observations of CH$_2$OHCHO $1_{1,0}-1_{0,1}$ (13476.995 MHz, E$_u$=1.2 K) and HOCH$_2$CH$_2$OH $2_{0,2}(v=0)- 1_{0,1}(v=1)$ (13380.638 MHz, E$_u$=1.5 K) in Sgr B2 were carried out during March and November 2016. The Digital backend system (DIBAS) (Li et al. 2016) was used, with a total bandwidth of 1.2 GHz, and a velocity resolution of 2.0 km s$^{-1}$ at a frequency of 13.5 GHz. The half power beam width is $\sim$ 84\arcsec\ at 13.5 GHz. The point-by-point map was obtained around Sgr B2(N) (RA (J2000) = 17:47:19.8, Dec (J2000) = -28:22:17.0) with a grid of 60\arcsec. About 100 points are observed in total. Spectra were taken in the position-switching mode with the reference position 60\arcmin\ east in azimuth with respect to the ON-source position. A single scan consisting of 2 minutes in the ON-source position followed by 2 minutes in the OFF-source position. The typical on-source integration time for a sampling point was 48 minutes. The rms noise level is about 4-8 mK under a velocity resolution of 2.0 km s$^{-1}$. The pointing accuracy is better than 12\arcsec. The wide frequency coverage allows for simultaneous observations of several emission lines, including CH$_2$OHCHO $1_{1,0}-1_{0,1}$, HOCH$_2$CH$_2$OH $2_{0,2}(v=0)- 1_{0,1}(v=1)$, HC$_5$N 5-4 (13313.334 MHz), H78$\alpha$ (13595.49 MHz), and absorption lines including H$_2$CO $2_{1,1}-2_{1,2}$ (14488.48 MHz) and so on.

The amplitude calibration was done by injecting periodic noise, and the accuracy is better than 20\%. The system temperature was 40-80 K. The maximum elevation of Sgr B2 from TMRT observation is $\sim$30$^{\circ}$, and the measurements were made at elevations above 15$^{\circ}$. At such low elevations, the resulting antenna temperatures were scaled to main beam temperatures ($T_{MB}$) by using a main beam efficiency of 0.4. The data processing was conducted using \textbf{GILDAS} software package\footnote{\tt http://www.iram.fr/IRAMFR/GILDAS.}, including CLASS and GREG. Linear baseline subtractions were performed for most of the spectra, and a two- or three-order baseline subtraction was introduced when necessary. For each transition, the spectra of subscans, including two polarizations, were averaged to improve the signal-to-noise ratio.

\subsection{Line Identifications}

Observations have shown that Sgr B2 complex contains a number of young O-type stars, compact HII regions, and molecular clouds with complex structure and composition. Line identification of CH$_2$OHCHO $1_{1,0}-1_{0,1}$ and HOCH$_2$CH$_2$OH $2_{0,2}(v=0)- 1_{0,1}(v=1)$ is affected by contamination from radio recombination line (RRL) and emissions from other species. To identify CH$_2$OHCHO $1_{1,0}-1_{0,1}$, we compare line profiles of H78$\alpha$ and H$_2$CO $2_{1,1}-2_{1,2}$. Positions with strong H78$\alpha$ emission would have corresponding H123$\delta$ (13474.755 MHz) next to CH$_2$OHCHO. RRL emission is especially strong toward Sgr B2(N) and Sgr B2(M). Figure 1 shows spectra toward Sgr B2(N) and Sgr B2(M). The detection of CH$_2$OHCHO $1_{1,0}-1_{0,1}$ is secure for these two positions, though it is blended with H123$\delta$ for Sgr B2(M). Gaussian fitting was used to determine the line parameters. During Gaussian fitting process, two Gaussian components were used to fit the CH$_2$OHCHO and the H123$\delta$ emission, while the third Gaussian was used to fit the He123$\delta$ emission if necessary. Through such process, the affect of H123$\delta$ on determination of line parameters could be fully eliminated, while there is no H123$\delta$ emission at all in cold regions. Thus this will not lead to an overestimate of the mass.

We found that $^{17}$OH (N=3-3, J+1/2=7/2-7/2, P=-1-1, F$_1$=3-4, F+1/2=3/2-3/2, 13380.4312 MHz, E$_u$= 290.3 K) is near the rest frequency of HOCH$_2$CH$_2$OH. The data for $^{17}$OH are provided by the JPL catalog (Pickett et al. 1998) through the Splatalogue interface. 
 We checked similar transitions of OH at 13441.4173 MHz and detected absorption lines in only a few points such as Sgr B2(N) and Sgr B2(M). As is shown in Figure 2, the peak intensities of OH are -0.03 K and -0.15 K for Sgr B2(N) and Sgr B2(M), respectively. Since $^{16}$O/$^{17}$O is observed to be up to 1000 in Sgr B2 (Polehampton et al. 2005; Zhang et al. 2015), $^{17}$OH should appear as absorption line with very low peak intensity, which could be ignored safely.

There is an emission feature next to CH$_2$OHCHO line at some positions, where there is no H78$\alpha$ emission. H$_2$CO is thought to be much more abundant and widespread than these two COMs. If this emission feature is another velocity component of CH$_2$OHCHO emission, it should have corresponding H$_2$CO absorption within the similar velocity range. Figure 3 (upper panel) shows spectra toward positions with offset (420\arcsec, -240\arcsec) away from Sgr B2(N), with LSR velocity calculated for the rest frequency of CH$_2$OHCHO $1_{1,0}-1_{0,1}$ (black), HOCH$_2$CH$_2$OH $2_{0,2}(v=0)- 1_{0,1}(v=1)$ (red), H78$\alpha$ (green) and H$_2$CO $2_{1,1}-2_{1,2}$ (blue). The LSR velocity range for H$_2$CO absorption line is 10$\sim$70 km s$^{-1}$. Two emission features were seen near the rest frequency of CH$_2$OHCHO $1_{1,0}-1_{0,1}$, with frequency separation of $\sim$2.1 MHz for the emission peaks. As shown in {\bf Figure 3a}, one emission feature has a velocity range of 10$\sim$50 km s$^{-1}$ while calculating for the rest frequency of CH$_2$OHCHO $1_{1,0}-1_{0,1}$, which is within velocity range of H$_2$CO absorption. This emission feature was attributed to CH$_2$OHCHO $1_{1,0}-1_{0,1}$. Another emission feature has a velocity range of 60$\sim$100 km s$^{-1}$, which is outside the velocity range of H$_2$CO absorption. We attributed this emission feature to CH$_3$CH$_2$CHO $2_{1,1}-2_{0,2}$ (13474.875 MHz, E$_u$=2.2 K), which has been detected in absorption toward Sgr B2(N) (Hollis et al. 2004). The data for CH$_3$CH$_2$CHO were taken from the Spectral Line Atlas of Interstellar Molecules (SLAIM) that is available through the Splatalogue interface\footnote{\tt www.splatalogue.net.}.

\section{RESULTS}
\label{result}

\subsection{Observing results}

Figure 3b and 4 show profile maps of CH$_2$OHCHO (black), HOCH$_2$CH$_2$OH (red), H78$\alpha$ (green) and H$_2$CO (blue) for the eastern region of Sgr B2 complex and the central region of Sgr B2, respectively, with a smoothed velocity resolution of 4 km s$^{-1}$. The shaded velocity range shows the emission channels of molecular lines. Figure 4 (line 4, row 3) shows spectra of CH$_2$OHCHO and HOCH$_2$CH$_2$OH for Sgr B2(M) with an offset (0, -60) relative to Sgr B2(N) observed at 13.5 GHz. The peak intensity of CH$_2$OHCHO and HOCH$_2$CH$_2$OH are 38 mK and 35 mK, respectively. Both are lower than those in Sgr B2(N), which are 50 mK and 41 mK, respectively. Thus the 100 GHz emission of these two molecules in Sgr B2(M) should be also lower than those in Sgr B2(N) with assumption of similar excitation temperatures. The 100 GHz emission of CH$_2$OHCHO ranges from 15-45 mK in Sgr B2(N) (Hollis et al. 2000; Halfen et al. 2006), and both CH$_2$OHCHO and HOCH$_2$CH$_2$OH were weakly detected in IRAM 30m line survey of Sgr B2(N) (Belloche et al. 2013). It is no wonder that both CH$_2$OHCHO and HOCH$_2$CH$_2$OH were not detected in the IRAM 30m line survey of Sgr B2(M) with a RMS noise level of 15-45 mK at 100 GHz (Belloche et al. 2013).  

The kinematics of the Sgr B2 complex is very complicated. At the eastern region of Sgr B2 complex, the CH$_2$OHCHO and HOCH$_2$CH$_2$OH emission lines cover the velocity range of 20 to 60 km s$^{-1}$, peaking at 30-40 km s$^{-1}$ (Figure 3b), while these two emission lines span the velocity range of 50 to 90 km s$^{-1}$, peaking at about +64 km s$^{-1}$ at the central region of Sgr B2 complex(Figure 4). The linewidth derived from Gaussion fitting generally range from 20 to 40 km s$^{-1}$, and up to 50 km s$^{-1}$ at some positions. The line profiles of these two emission lines are similar to those of H$_2$CO, implying that these emissions come from the quiescent gas. Since the emission is weak, the SNR is not good enough for discussion of the kinematics in detail. 

Figure 5 displays contour maps of glycolaldehyde, ethylene glycol and HC$_5$N overlaid on the H78$\alpha$ emission in color scale, which demonstrate the spatial coincidence among glycolaldehyde, ethylene glycol and HC$_5$N. Two main components were distinctly observed in the contour maps: the southeastern component spans a velocity range of 20 - 60 km s$^{-1}$, and the northwestern component spans a velocity range of 50 - 90 km s$^{-1}$. The northwestern component is associated with Sgr B2(N), while the southeastern component is not associated with any known HII region. The Sgr B2 complex consists of a 5\arcmin\ long north-south chain of bright clumps, surrounded by a halo of fainter ones, filaments, and shells extending over nearly 20\arcmin\ in R.A. and 10\arcmin\ in declination (Bally et al. 2010), thus the southeastern component is also part of Sgr B2 complex. This region seems to be colder than the central region of Sgr B2 complex, since no RRL emission was detected in our observations. The Bolocam Galactic Plane Survey (BGPS) 1.1 mm data shows that the southeastern component is associated with clump with H$_2$ column density of about $10^{22}$ cm$^{-2}$ (Figure 17 in Bally et al. 2010). Several peaks were observed in both the CH$_2$OHCHO and HOCH$_2$CH$_2$OH maps. Though the emission peaks of these two molecules occur at different positions, the overall emission regions are not spatially distinct. These maps show that the CH$_2$OHCHO and HOCH$_2$CH$_2$OH emissions span a range of approximately 15 arc-minutes, corresponding to a physical size of $\sim$ 36 pc adopting a distance of 8.34 kpc from the Sun (Reid et al. 2014). The HOCH$_2$CH$_2$OH emission appears to be more extended than that of CH$_2$OHCHO, with more extended emissions detected above 5$\sigma$ levels. The extension of these two molecules in Sgr B2 is more than 700 times greater than that in spiral arm clouds.

Complex aldehydes and alcohols, including glycolaldehyde and ethylene glycol, have been detected in three typical molecular clouds in the Central Molecular Zone (CMZ) (Requena-Torres et al. 2008). One of these clouds, G+0.693-0.03 is (30\arcsec, 50\arcsec) away from Sgr B2(N). Our sampling interval is 60\arcsec, thus it is in the middle of (0, 60\arcsec) and (60\arcsec, 60\arcsec), and is covered by our observations. As is shown in Figure 4, results of (0, 60\arcsec) and (60\arcsec, 60\arcsec) are consistent with that of G+0.693-0.03 (Requena-Torres et al. 2008). The chiral molecule propylene oxide was also discovered in a cold, extended molecular shell around the massive protostellar clusters in Sgr B2 but not in hot cores (McGuire et al. 2016). However, because of its overall large extent, the TMRT observations presented here provide direct evidence for the existence of a huge mass of organic molecules in our Galactic center region. Importantly, results presented here also show that most of the emission is not associated with star formation.

\subsection{Column densities}

With Green Bank Telescope (GBT) observations, Hollis et al. (2004) obtained an excitation temperature T$_{ex}$ = 8 K toward the position of Sgr B2(N). Assuming the excitation temperature is typical for the Sgr B2 positions in which glycolaldehyde was detected by TMRT, we calculate the total column densities of CH$_2$OHCHO for the i, j-th grid in the position-position space $N_{ij}$ with the expression reported by Hollis et al. (2004):
\begin{eqnarray}
N_{ij}=\frac{3kQe^{E_u/kT_S}}{8 \pi^3 \nu S \mu^2 } \times \frac {\frac{1}{2} \sqrt{\frac{\pi}{ln 2}} \frac{\triangle T_A^* \triangle V}{\eta_B}} {1-\frac{e^{h\nu/kT_S}-1}{e^{h\nu/KT_{bg}}-1}} ,
\end{eqnarray}
in which $k$ is the Boltzmann constant in erg K$^{-1}$, $\frac{\triangle T_A^* \triangle V}{\eta_B}$ is the observed line integrated intensity in K Km s$^{-1}$, $\nu$ is the frequency of the transition in Hz, and $S\mu^2$ is the product of the total torsion-rotational line strength and the square of the electric dipole moment. $T_{S}$ and $T_{bg}$ (=2.73 K) are the excitation temperature and background brightness temperature, respectively. $E_u/k$ is the upper level energy in K. The partition function, $Q$, was estimated by fitting the partition function at different temperatures given in CDMS (M$\ddot{u}$ller et al. 2005). Values of $E_u/k$ and $S\mu^2$ are also taken from CDMS.
The column densities range from $1.3 \times 10^{14}$ to $4.6 \times 10^{14}$ cm$^{-2}$ depending on the position.
Data points with integrated intensities above 5$\sigma$ levels were then used for mass calculation. The total mass of CH$_2$OHCHO is computed  using the relation:
\begin{eqnarray}
M=\Sigma N_{ij}\delta x_i \delta y_j m_{molecule},
\end{eqnarray}
in which $N_{ij}$ is the column density of CH$_2$OHCHO at the i, j-th grid in the position-position space. $m_{molecule}$ is the mass of molecule. The spacings of $\delta x_i$ and $\delta y_j$ are the sizes of grid in the RA and Dec directions, respectively. Adopting a distance of 8.34 kpc, $\delta x_i$ = $\delta y_j$ = 60 \arcsec $\times$ 8.34 kpc  $\times \frac{3.1 \times 10^{18} cm}{206265} =7.3 \times 10^{18}$ cm. The total mass of CH$_2$OHCHO is estimated to be about $6.1\times10^{31}$ g, corresponding to $1.0\times10^4 M_{Earth}$. 
The column densities of HOCH$_2$CH$_2$OH were also calculated using equation (1) with an excitation temperature of 8 K. In the equation, the partial function $Q(T_S)=133T_S^{1.5}$, was estimated by fitting the partition function at different temperatures given in CDMS (M$\ddot{u}$ller et al. 2005). $E_u/k = 1.47 K$, the upper level energy in $K$, was taken from CDMS through Splatalogue. The product of the transition line strength times the square of coupled dipole momentum, $S\mu^2=61.2$, was also taken from CDMS. The column densities of ethylene glycol range from $3.4 \times10^{14}$ to $13\times10^{14}$ cm$^{-2}$ for different points, implying a total mass of approximately $4.2\times10^4 M_{\rm Earth}$. Both the typical column density and total mass of HOCH$_2$CH$_2$OH are higher than those of CH$_2$OHCHO. With a T$_{ex}$ of 30 K, one can obtain a column density of $2.3\pm0.7 \times10^{15}$ cm$^{-2}$ for CH$_2$OHCHO at the position of Sgr B2(N), suggesting that a high excitation temperature of 30 K would increase column densities and total mass of glycolaldehyde by a factor of about 4. This column density of Sgr B2(N) does not conflict with that at 3-mm wavelength (Belloche et al. 2013), which is calculated to be $1.8\times10^{15}$ cm$^{-2}$. Similarly, for ethylene glycol, we note that an excitation of 30 K would increase column density by a factor of about 3. The result is in agreement with that at 3-mm wavelength (Belloche et al. 2013).

\section{DISCUSSIONS}
\label{discussion}

 \subsection{Abundance}

We made use of the H$_2$ column density obtained with BGPS 1.1 mm data (Bally et al. 2010) for abundance estimation of glycolaldehyde and ethylene glycol. For each position, we searched for nearest clump identified with BGPS data, and obtained the corresponding hydrogen column density. The H$_2$ column densities range from $1.4\times10^{22}$ cm$^{-2}$ to $1.0\times10^{25}$ cm$^{-2}$ in our observing region, and peak toward Sgr B2(N) and Sgr B2(M). The abundance of CH$_2$OHCHO is $\sim2\times10^{-10}$ toward Sgr B2(N) and Sgr B2(M), and is $\sim2\times10^{-9}$ toward the southeastern region of Sgr B2 complex, which is colder than the central star-forming region. Similarly, the abundance of HOCH$_2$CH$_2$OH is $\sim2\times10^{-10}$ toward Sgr B2(N) and Sgr B2(M), and is up to $\sim2\times10^{-8}$ toward the colder region. The abundance of these two molecules are lower than those in Requena-Torres et al. (2008), which is caused by different hydrogen column densities adopted. Fortunately this {\bf does} not affect the overall trend. Figure 6 shows variation of CH$_2$OHCHO (circle) and HOCH$_2$CH$_2$OH (square)  abundance with the R.A. offset away from Sgr B2(N). Generally, regions with small offset ($-150\arcsec < offset < 200\arcsec$) are associated with star-formation, while regions with large absolute values of offset are away from star-forming region. We could see from Figure 6 that the abundance of these two molecules generally decrease toward positions with small offset, implying that the abundance of these two molecules decrease from cold to hot region. Note that points with small offset but high abundance ( $>10^{-8}$) come from regions on the north of Sgr B2(N), where no RRL emission was detected. This may suggest that these two molecules are produced under low temperature, and are destroyed in star-forming process. The implication of these results for formation mechanism of these two molecules will be discussed in Section 4.3. 

\subsection{Abundance Ratio of Ethylene glycol to glycolaldehyde}

Observations in different star-forming regions indicate that the abundance ratio of ethylene glycol to glycolaldehyde (hereafter EG/GA) range from 1 to $>15$ (J${\o}$rgensen et al. 2016; Coutens et al. 2015; Fuente et al. 2014; Biver et al. 2015; Rivilla et al. 2017; Lykke et al. 2015). Comparing different star-forming regions, Rivilla et al. (2017) found evidence for the increase of abundance ratio with the luminosity of the source. Figure 7 shows variation of EG/GA abundance ratio with R.A. offset for positions with both of these two molecules detected. The EG/GA ratio is up to $\sim$6 toward the cold region, while it is $\sim$1 toward the central regions with active star-formation activity. Requena-Torres et al. (2008) observed glycolaldehyde and ethylene glycol toward three clouds in the GC, including 20 km s$^{-1}$ cloud, 50 km s$^{-1}$ cloud and G+0.693-0.03, which is (30\arcsec, 50\arcsec) away from Sgr B2(N). They obtained EG/GA ratios of 1.2, 1.6 and 1.3 for these three clouds, which are similar to those found for Sgr B2(N) and Sgr B2(M). All of these three clouds are associated with star-formation activity (Lu et al. 2015; Tsuboi et al. 2015), thus their result also supports low EG/GA ratio in star-forming regions in the GC. These results seem to be in contradiction with those found in the Galactic disk sources (Rivilla et al. 2017), suggesting that the formation mechanism of these two species might be different in Galactic disk and Galactic center sources. However, high angular resolution and more sensitive maps are still needed to identify and study these COM clumps in detail. 

\subsection{Formation Mechanism}

Due to the inefficiency of gas-phase reactions for saturated complex species (Geppert et al. 2006), the grain-surface mechanism is regarded to be important for the formation of complex organic molecules (COMs) including glycolaldehyde and ethylene glycol. Many grain-surface formation routes have been proposed (Bennett et al. 2007; Beltran et al. 2009; Woods et al. 2012; Garrod et al. 2008), and some of which involve thermal and energetic processes (Butscher et al. 2016). In this kind of models, the warm-up of the hot core is thought to be crucial for the formation of COMs, allowing the more strongly bound radicals to become mobile on the grain surfaces (Garrod et al. 2008). Experimental studies also demonstrated that photochemistry in CH$_3$OH-CO ice mixtures is efficient enough to explain the observed abundance of glycolaldehyde and ethylene glycol ($\ddot{O}$berg et al. 2009). This type of chemistry appears to be scapable of reproducing the high degree of complexity seen around protostars. Since the cosmic ray ionization rate is high in the CMZ (e.g. Barnes et al. 2017), the cosmic ray induced UV field is expected to be high everywhere, not only towards the star-forming regions. This UV field could also play a role in the formation of COMs in Sgr B2, even in regions devoided of star formation activity.

Theoretical and experimental studies have shown that, along with the formation of H$_2$CO and CH$_3$OH, molecules with more than one carbon atom can be formed on interstellar grains at low temperature (Woods et al. 2013; Fedoseev et al. 2015; Butscher et al. 2015). Two different pathways appear most likely to occur, with the first mechanism involving the radial combination of HCO with CH$_2$OH to form glycolaldehyde and radical CH$_2$OH dimerization to form ethylene glycol. 
In the second mechanism, the key step is the recombination of two HCO radicals followed by the formation of a C-C bond. Sequential hydrogenation by two or four H atoms transforms HC(O)CHO into glycolaldehyde and ethylene glycol, respectively. The detection of widely distributed glycolaldehyde and ethylene glycol in Sgr B2 complex, as well as the decrease of abundance toward star-forming region, can be explained by such a low temperature process (Chuang et al. 2016). COMs like acetaldehyde, formic acid, ketene, and propyne have also been detected in a few prestellar cores (Jimenez et al. 2016; Lefloch et al. 2017) and cold dark cloud region (Taquet et al. 2017). We noted that simulations of a low temperature model with $T=12 K$ and $n(H)=10$ cm$^{-3}$ (see Figure 5c in Fedoseev et al. 2015) could produce the column densities of CH$_2$OHCHO and HOCH$_2$CH$_2$OH presented here. 

Chuang et al. (2017) carried out some laboratory experiments where they mixed low temperature conditions and UV irradiation. They found that glycolaldehyde and ethylene glycol were always formed at low temperature with and without UV-photon absorption (i.e. energetic processes). They also found that methyl formate was more sensitive to energetic processes than glycolaldehyde and ethylene glycol. As methyle formate could not form efficiently in the absence of energetic processes, one way to investigate the role of energetic processes here would be to search for methyl formate. If it is found to be abundant, it would mean that energetic processes in addition to the low temperature play a role. In contrast, a low abundance of methyl formate would mean that only the low temperature process is important in the Galactic center.

Unfortunately, the formation routes of these two molecules are far from solved. Recent quantum chemical computations indicate that acetaldehyde cannot be synthesized by the CH$_3 +$ HCO coupling on the icy grains, suggesting that formation of some COMs on the grain surfaces might be unlikely (Erique-Romero et al. 2016). In fact, COMs like methyl formate and dimethyl ether have been proposed to form via gas-phase reactions from simpler precursors that were formed on grain surfaces and then ejected into the gas via efficient reactive desorption (Vasyunin et al. 2013; Balucani et al. 2015). The observed similar spatial distributions of glycolaldehyde, ethylene glycol and HC$_5$N, which is formed in the gas phase rather than in grain ices (Chapman et al. 2009), raises the possibility of gas-phase reactions as a complementary to the surface formation route of glycolaldehyde and ethylene glycol. However, there are very limited studies on gas-phase chemistry of glycolaldehyde and ethylene glycol (Halfen et al. 2006; Woods et al. 2012), and such possibilities need to be further investigated.

We searched for low-energy transitions of glycolaldehyde and ethylene glycol at 13.5 GHz towards IRAS 16293-2422, G31.41+0.31, G34.3+0.2 and NGC 7129 with TMRT but did not detect any emissions above 10 mK ($\sigma \sim$ 3 mK). Note that in a recent single-dish line survey of Orion KL at a frequency of 22 GHz (Gong et al. 2015), which covered HOCH$_2$CH$_2$OH $3_{0,3}(v=0)-2_{0,2}(v=1)$ (Requena-Torres et al. 2008), no emission was detected either. The absence of an extended, large-scale emission for these two molecules might result from the lack of a large-scale desorption mechanism to release them from the ices into the gas phase. In contrast, at the center of our Galaxy, prebiotic molecules preserved in icy mantles can be efficiently desorbed into the gas phase as a result of physical conditions specific to the Galactic center, such as high gas temperatures (Ginsburg et al. 2016), large-scale shocks (Requena-Torres et al. 2008; Menten 2011), and strongly enhanced cosmic rays. According to chemical modeling done by Coutens et al. (2016), the cyclic grain mantle explosions or very frequent shocks could reproduce the widespread presence of methanol in the CMZ (Yusef-Zadeh et al. 2013). Recently An et al. (2017) reported discovery of high CH$_3$OH ice abundance toward a star in the CMZ, implying that gas-phase CH$_3$OH in the CMZ can be largely produced by desorption from icy grains. Since more complex ice features cannot be directly observed at present, and most constraints on the ice compositions come from millimeter observations of desorbed ice chemistry products ($\ddot{O}$berg et al. 2011), observations presented here may provide important information on ice chemistry and imply the existence of abundant prebiotic molecules in ices.

 Widespread gas-phase prebiotic molecules may also exist in the nuclei of nearby galaxies, such as NGC 253, and in more distant starburst galaxies. Future interferometers operating at centimeter wavelengths, such as the Next Generation Very Large Array (ngVLA) and SKA, have the potential to probe such emissions to investigate whether the chemistry of CMZ clouds resembles that of starburst galaxies, and whether CMZ clouds could serve as a template for the nuclei of starburst galaxies in the nearby and distant universe (Kauffmann et al. 2016).

\section{CONCLUSIONS}
\label{conclusion}

    Our TMRT observations show that the spatial distribution of CH$_2$OHCHO and HOCH$_2$CH$_2$OH extend over 15 arc-minutes in Sgr B2, corresponding to a physical size of approximately 36 pc. The extension of these two molecules in Sgr B2 is more than 700 times greater than in spiral arm clouds. The HOCH$_2$CH$_2$OH emission appears to be more extended than that of CH$_2$OHCHO. The overall mass of glycolaldehyde is approximately $6.1\times 10^{31}$ g, roughly $10^4$M$_{\rm Earth}$. The TMRT observations provide direct evidence for the existence of a enormous mass of prebiotic molecules in our Galactic center regions. Results presented here show that most of the emission is not associated with star formation. The abundance of glycolaldehyde and ethylene glycol was found to decrease from the cold region to the central region associated with star-formation activity. The abundance ratio of ethylene glycol to glycolaldehyde also decreases toward central regions associated with star-formation activity, which seems to differ from observations in Galactic disk sources. The widely distributed glycolaldehyde and ethylene glycol, as well as the abundance decrease toward the star-forming region,a can be explained by a low temperature process. Future observations of methyl formate are expected to investigate whether energetic processes also play a role in producing COMs in the Galactic center.

\acknowledgements The authors are grateful to the anonymous referee for the constructive suggestions and very helpful comments on the paper. This study is based on observations carried out with the TMRT telescope. JL would like to than Dr Zhi-Yu Zhang for help in plot of profile maps. This work was supported in part by the National Natural Science Foundation of China (11590780, 11590784, U1431125 and 11773054), the Knowledge Innovation Program of the Chinese Academy of Sciences (Grant No. KJCX1-YW-18), the Scientific Program of Shanghai Municipality (08DZ1160100), ``CAS Interdisciplinary Innovation Team", ``the International Partnership Program" No.114A11KYSB20160008 and Key Laboratory for Radio Astronomy, CAS.

\begin{figure}
\begin{center}
\includegraphics[width=0.6\textwidth]{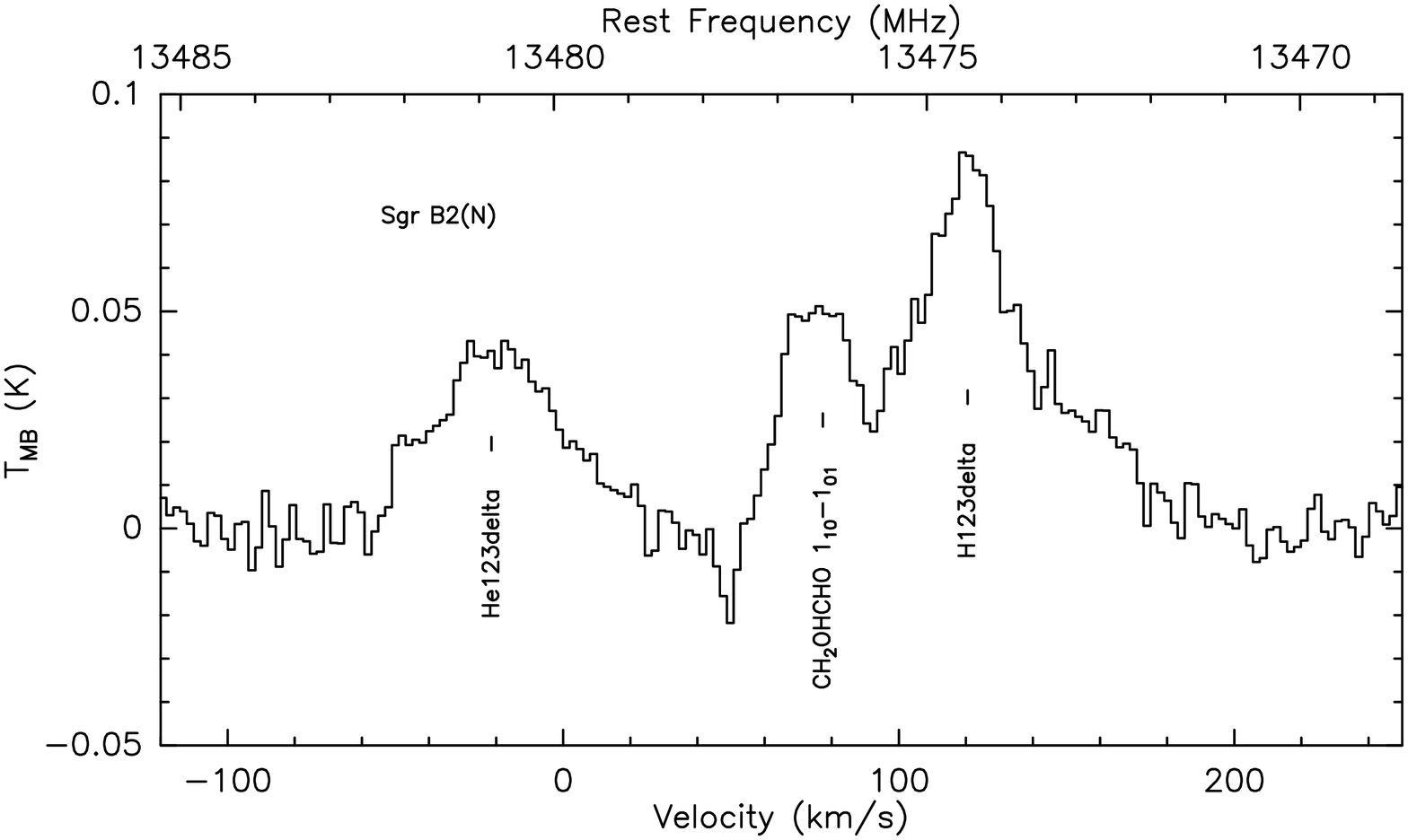}
 \includegraphics[width=0.6\textwidth]{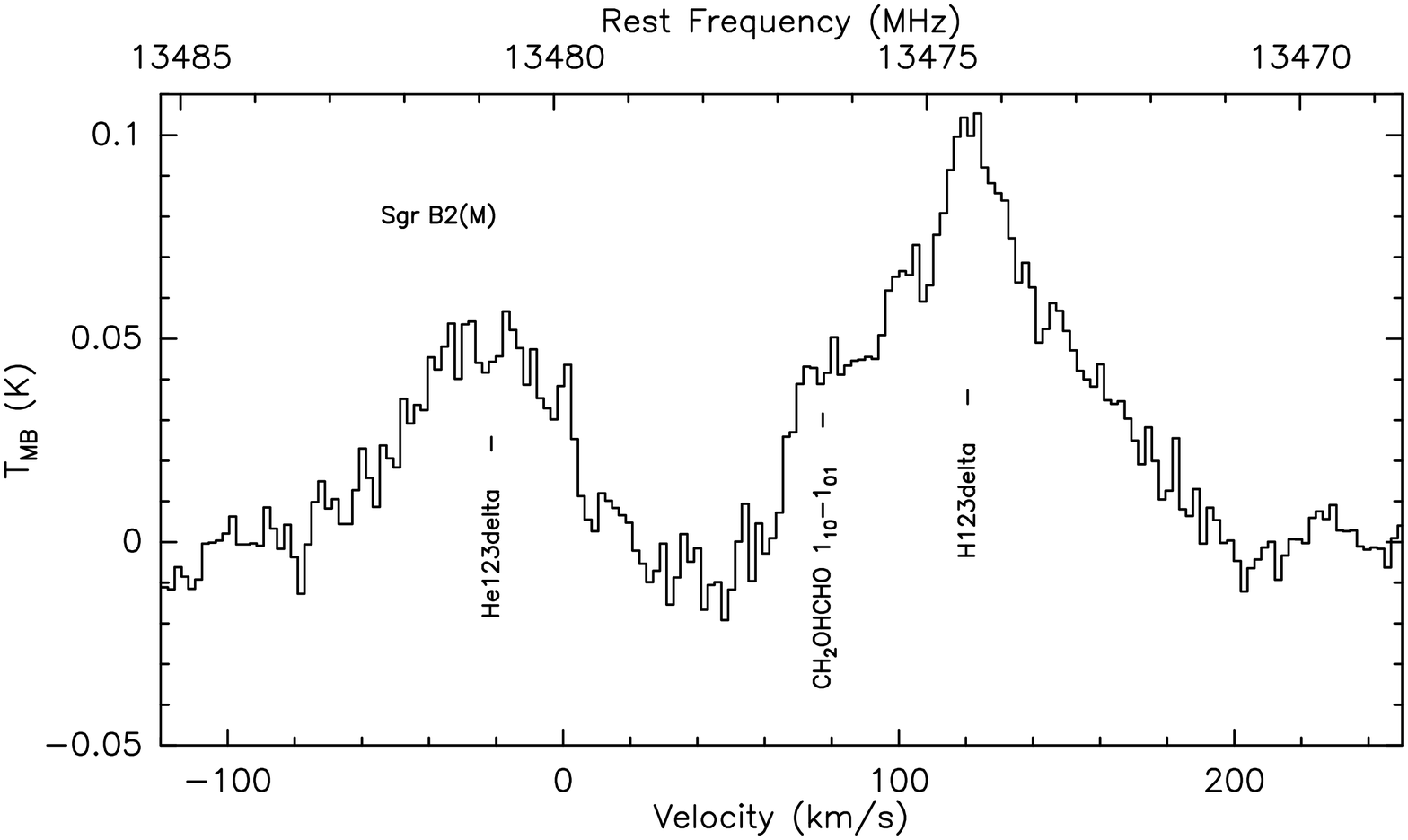} 
 \vspace{3cm}
\caption{Glycolaldehyde (CH$_2$OHCHO $1_{1,0}-1_{0,1}$) spectra toward Sgr B2(N) (upper panel) and Sgr B2(M) (lower panel) with a smoothed resolution of 4 km s$^{-1}$. The LSR velocity is calculated for the rest frequency of CH$_2$OHCHO $1_{1,0}-1_{0,1}$ at an assumed source velocity of +64 km s$^{-1}$. }
\end{center}
\end{figure}

\begin{figure}
\begin{center}
\includegraphics[width=0.6\textwidth]{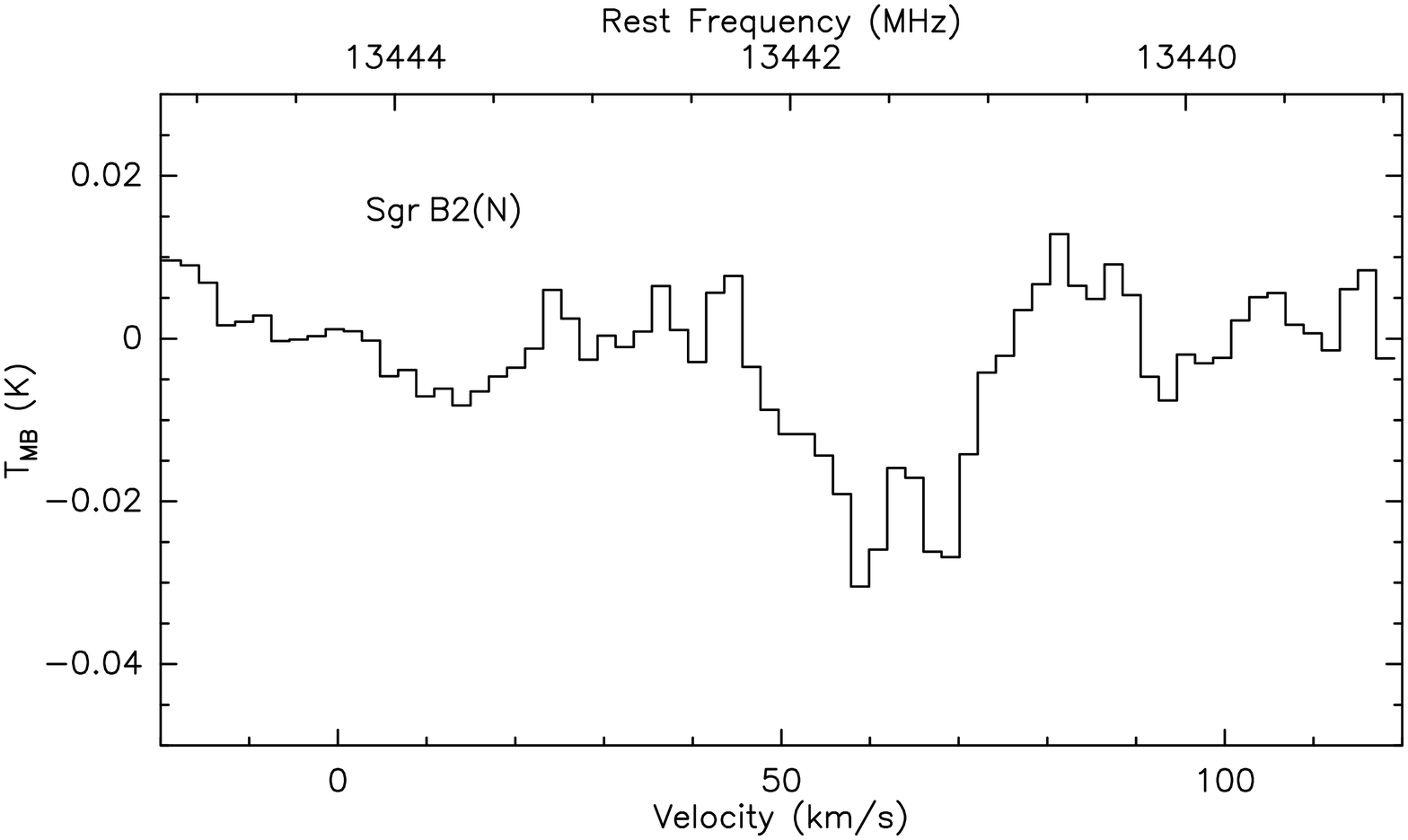}
 \includegraphics[width=0.6\textwidth]{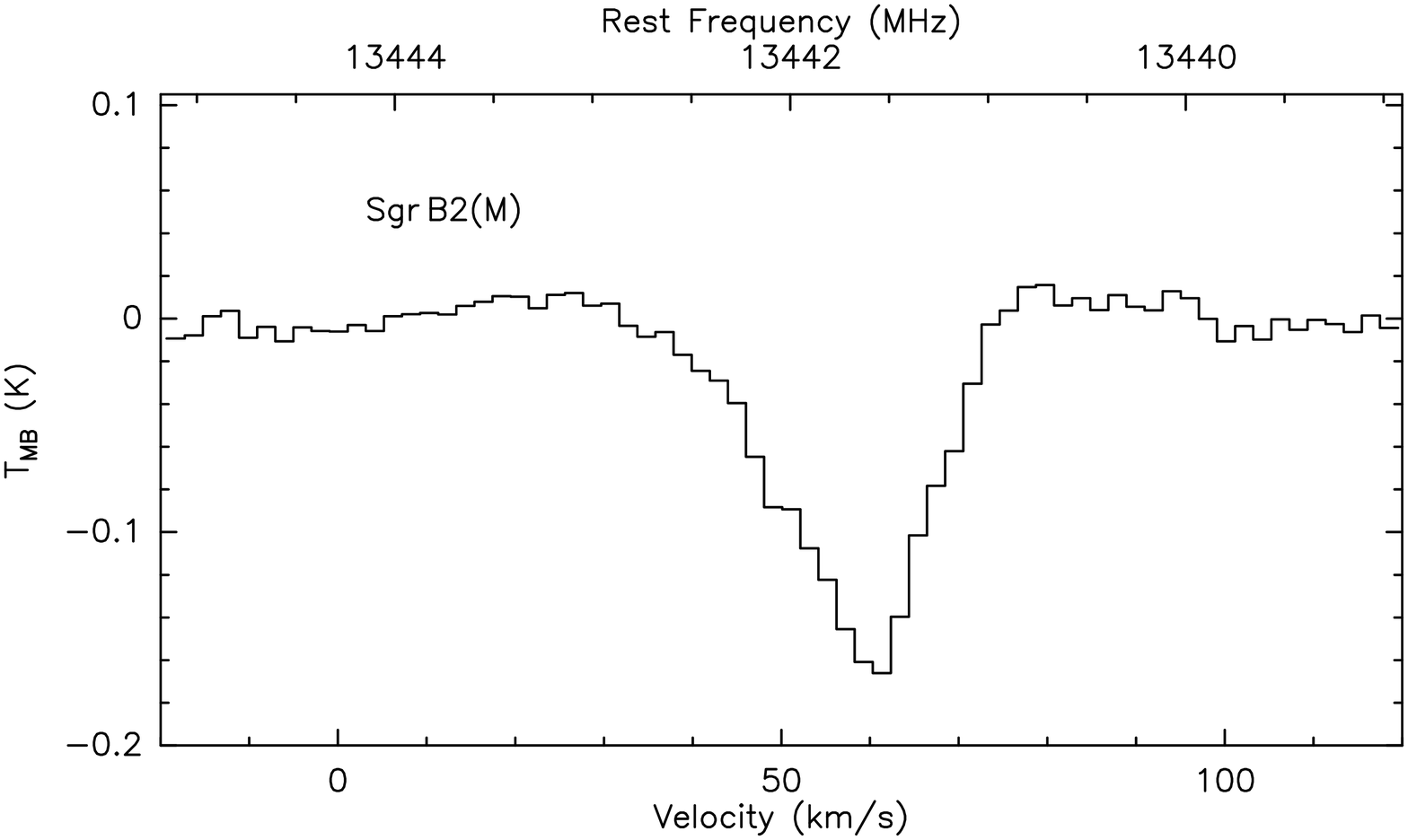} 
\vspace{3cm}
\caption{Spectra of OH (J=7/2, $\Omega$=3/2, F=$4^+-4^-$, 13441.4173 MHz) toward Sgr B2(N) (upper panel) and Sgr B2(M) (lower panel). }
\end{center}
\end{figure}

\begin{figure}
\begin{center}
\vspace{2cm}
 \includegraphics[width=0.4\textwidth]{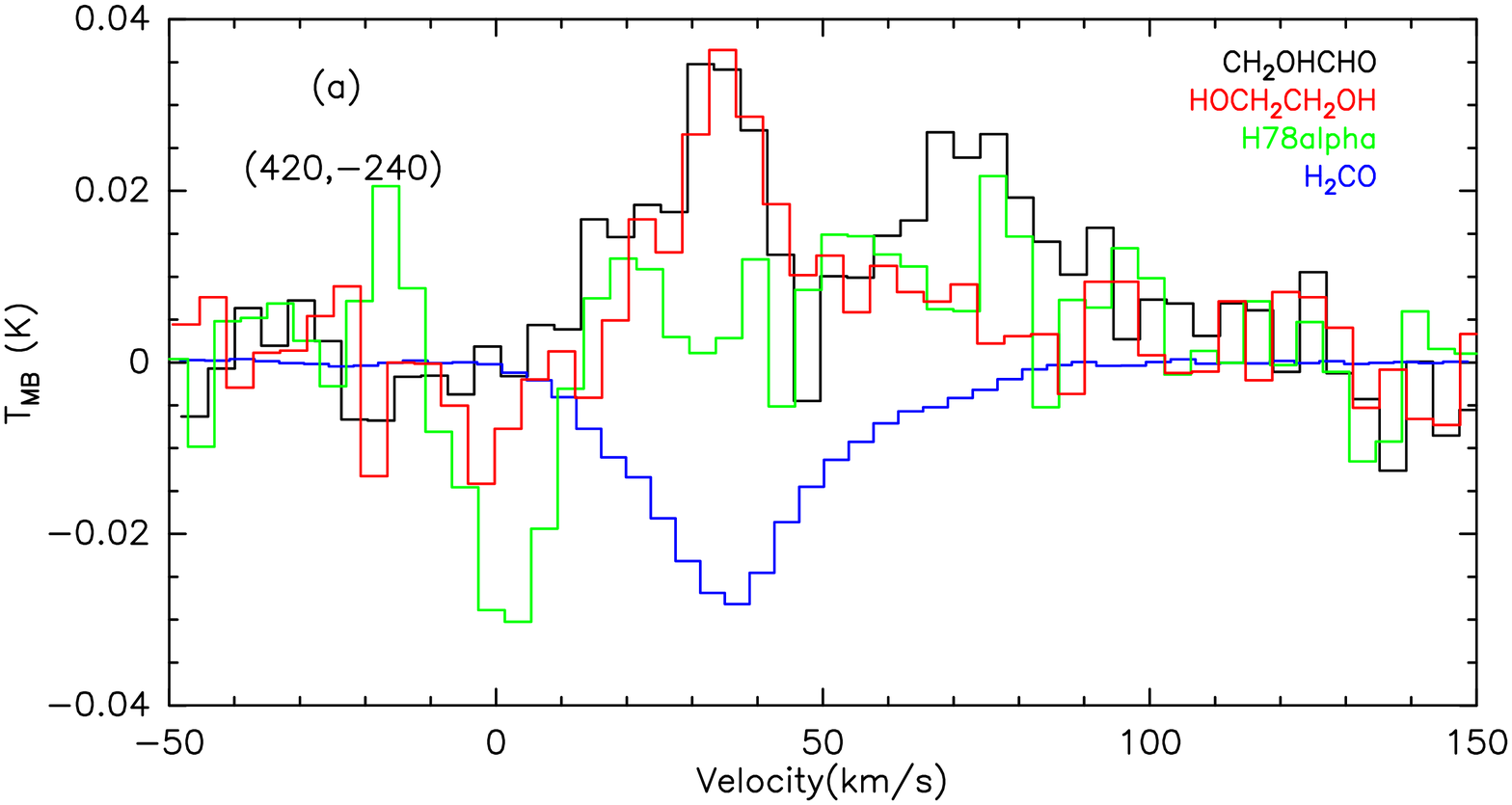} 
\vspace{10mm}
  \includegraphics[width=0.7\textwidth]{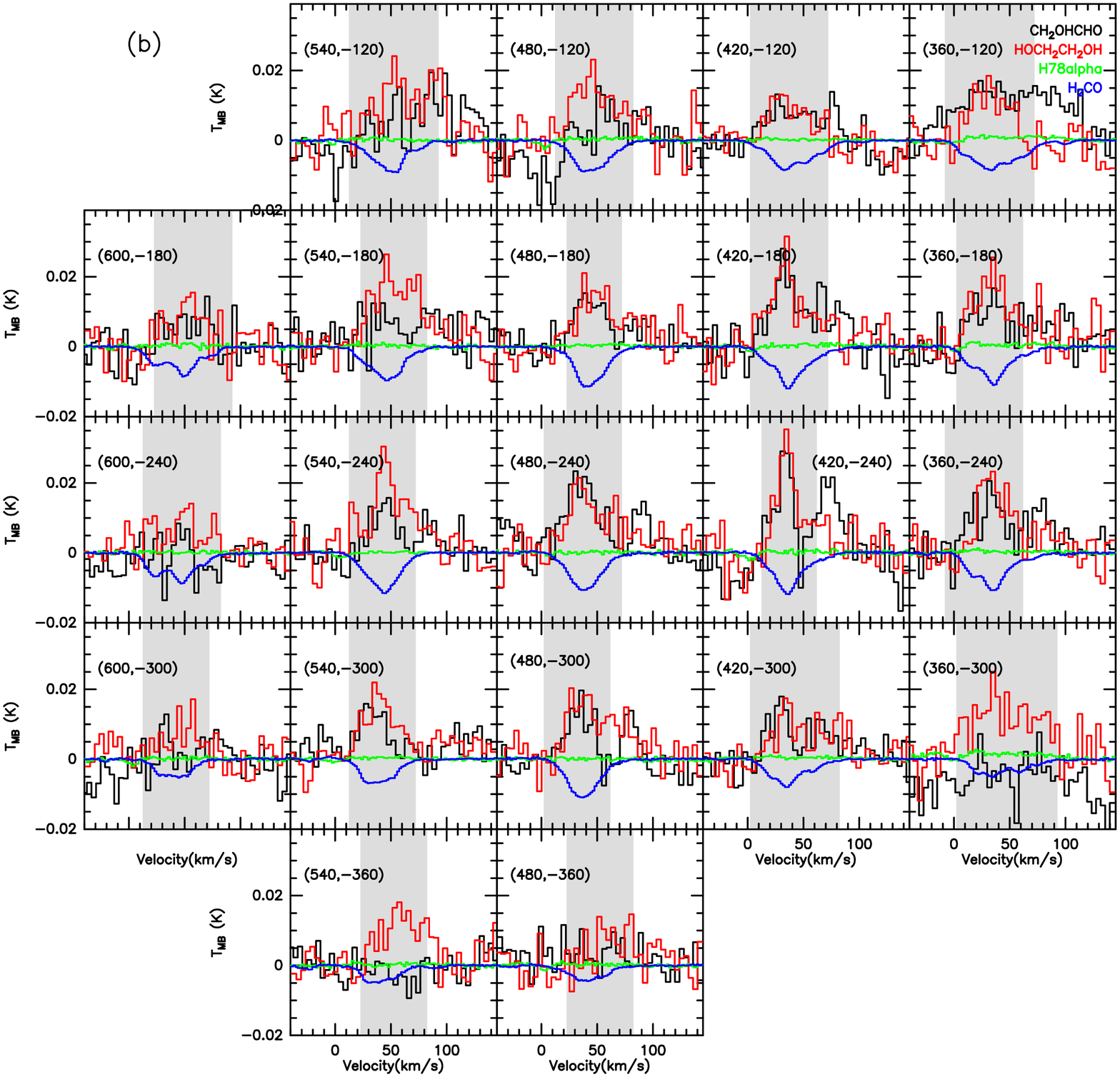} 
\caption{(a): Spectra of CH$_2$OHCHO $1_{1,0}-1_{0,1}$ (13476.995 MHz, black), HOCH$_2$CH$_2$OH $2_{0,2}(v=0)-1_{0,1}(v=1)$ (13380.638 MHz, red), H78$\alpha$ (13595.49 MHz, green) and H$_2$CO $2_{1,1}-2_{1,2}$ (14488.48 MHz, blue) with a smoothed resolution of 4 km s$^{-1}$ for positions with offset (420\arcsec, -240\arcsec) away from Sgr B2(N). The H$_2$CO is emission scaled down by 0.05 for plot. (b): Profile map of CH$_2$OHCHO $1_{1,0}-1_{0,1}$, HOCH$_2$CH$_2$OH $2_{0,2}(v=0)-1_{0,1}(v=1)$, H78$\alpha$ and H$_2$CO $2_{1,1}-2_{1,2}$ with a smoothed resolution of 4 km s$^{-1}$ for the western region of Sgr B2 complex. H$_2$CO emission is scaled down by 0.02 for plot. The grid spacing is 60 \arcsec. The offset relative to Sgr B2(N) (RA (J2000): 17:47:19.8, Dec (J2000): -28:22:17.0) is indicated in the figure. The shaded velocity range shows the emission channels of molecular lines. }
\end{center}
\end{figure}

\begin{figure}
\vspace{5cm}
\begin{center}
 \includegraphics[width=0.9\textwidth]{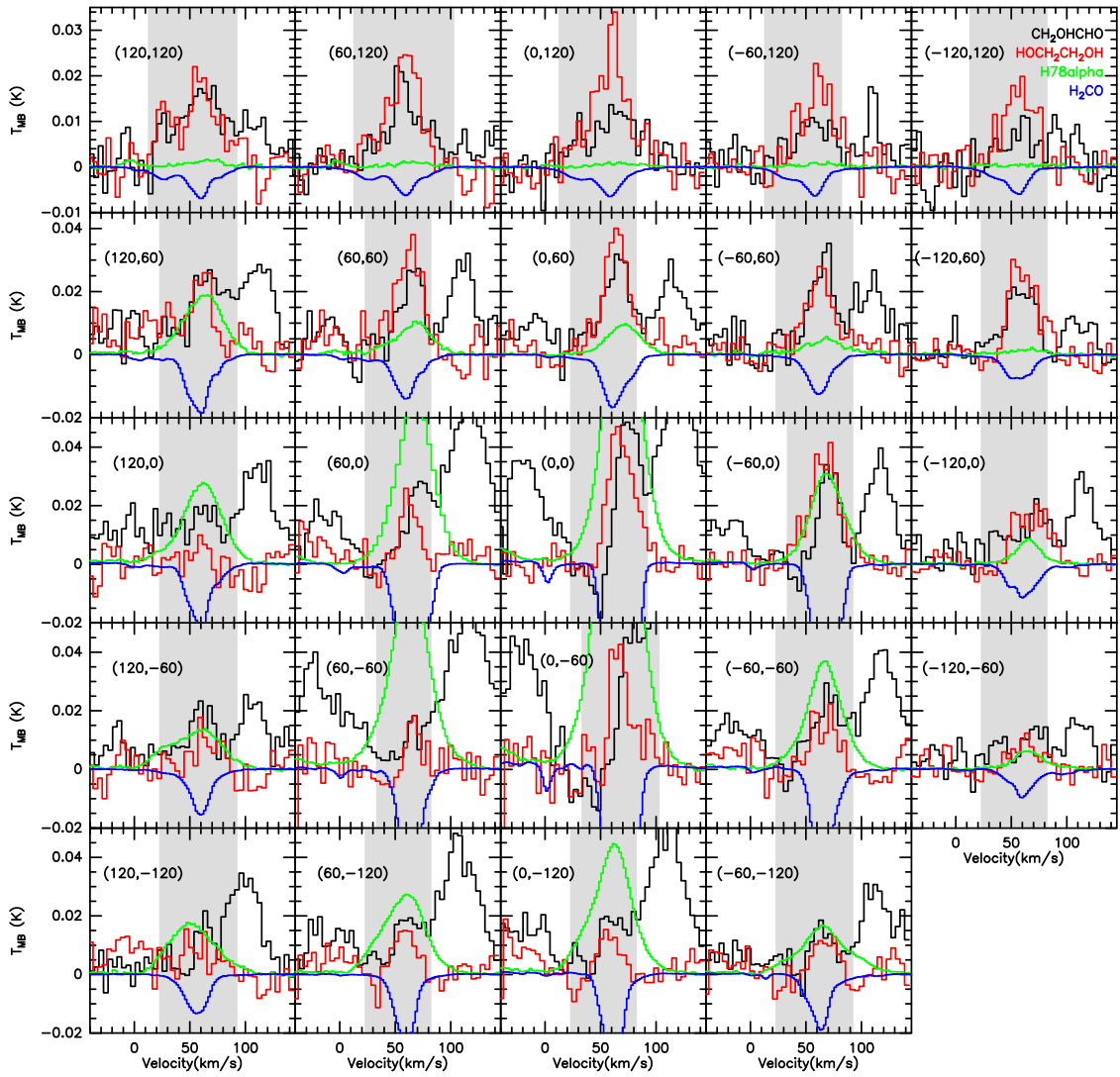}
\caption{Profile map of CH$_2$OHCHO $1_{1,0}-1_{0,1}$, HOCH$_2$CH$_2$OH $2_{0,2}(v=0)-1_{0,1}(v=1)$, H78$\alpha$ and H$_2$CO $2_{1,1}-2_{1,2}$ for the central region of Sgr B2 complex. The H78$\alpha$ emission is scaled down by 0.05, while H$_2$CO is scaled down by 0.01 for plot. Others are similar to Figure 3b.}
\end{center}
\end{figure}

\begin{figure}
\begin{center}
\vspace{-5mm}
\includegraphics[width=0.6\textwidth]{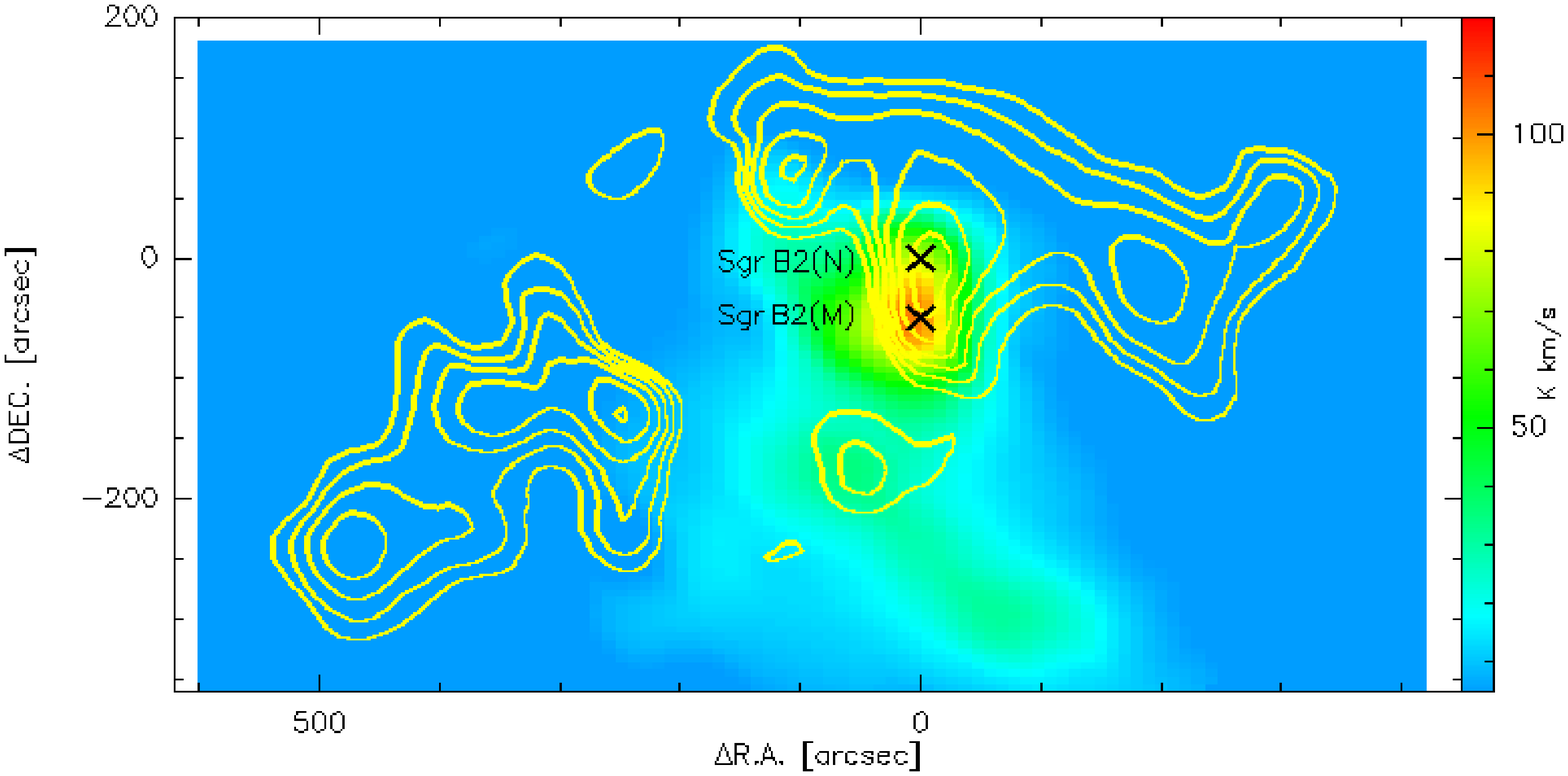} 
\includegraphics[width=0.6\textwidth]{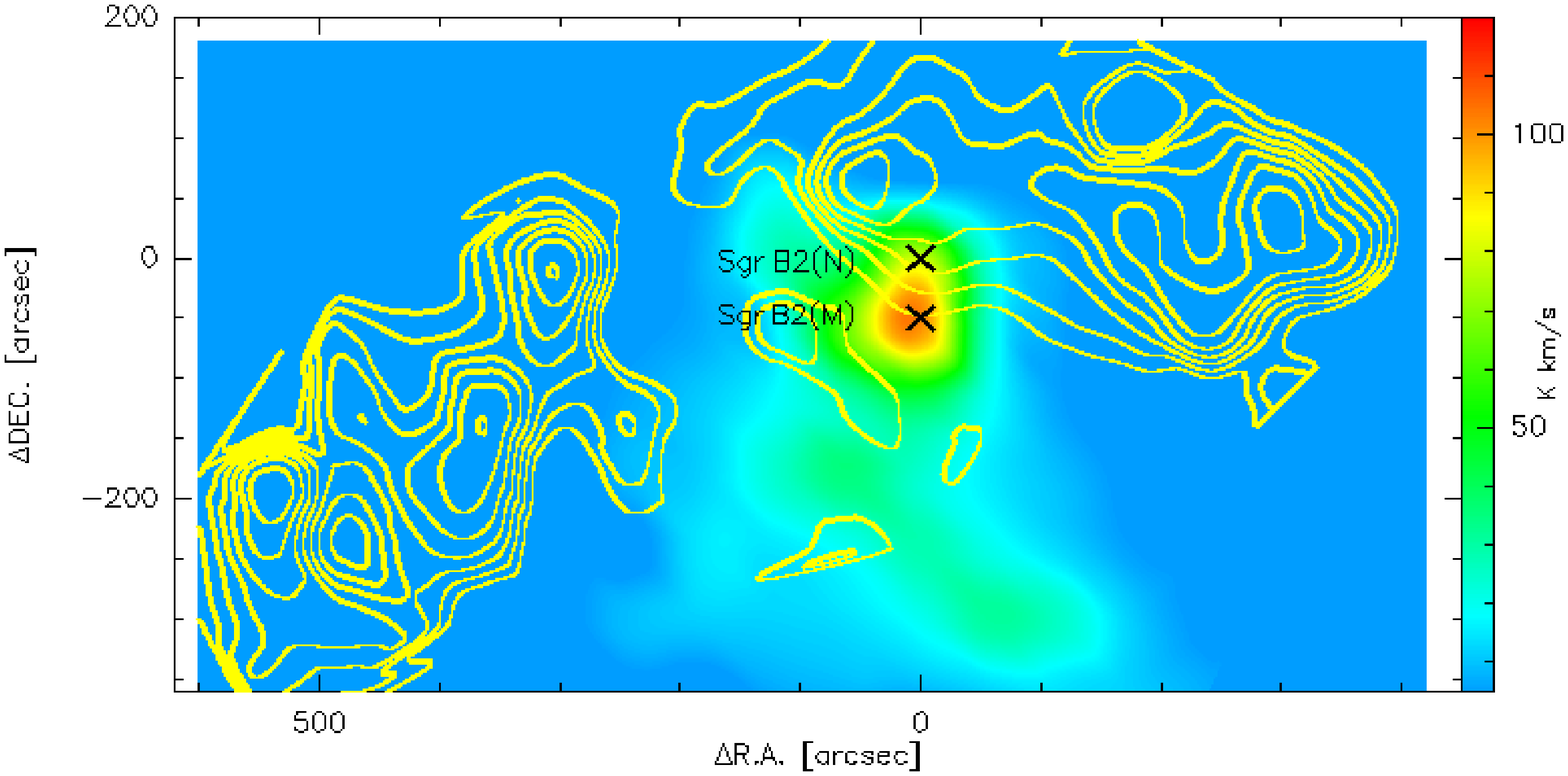} 
 \includegraphics[width=0.6\textwidth]{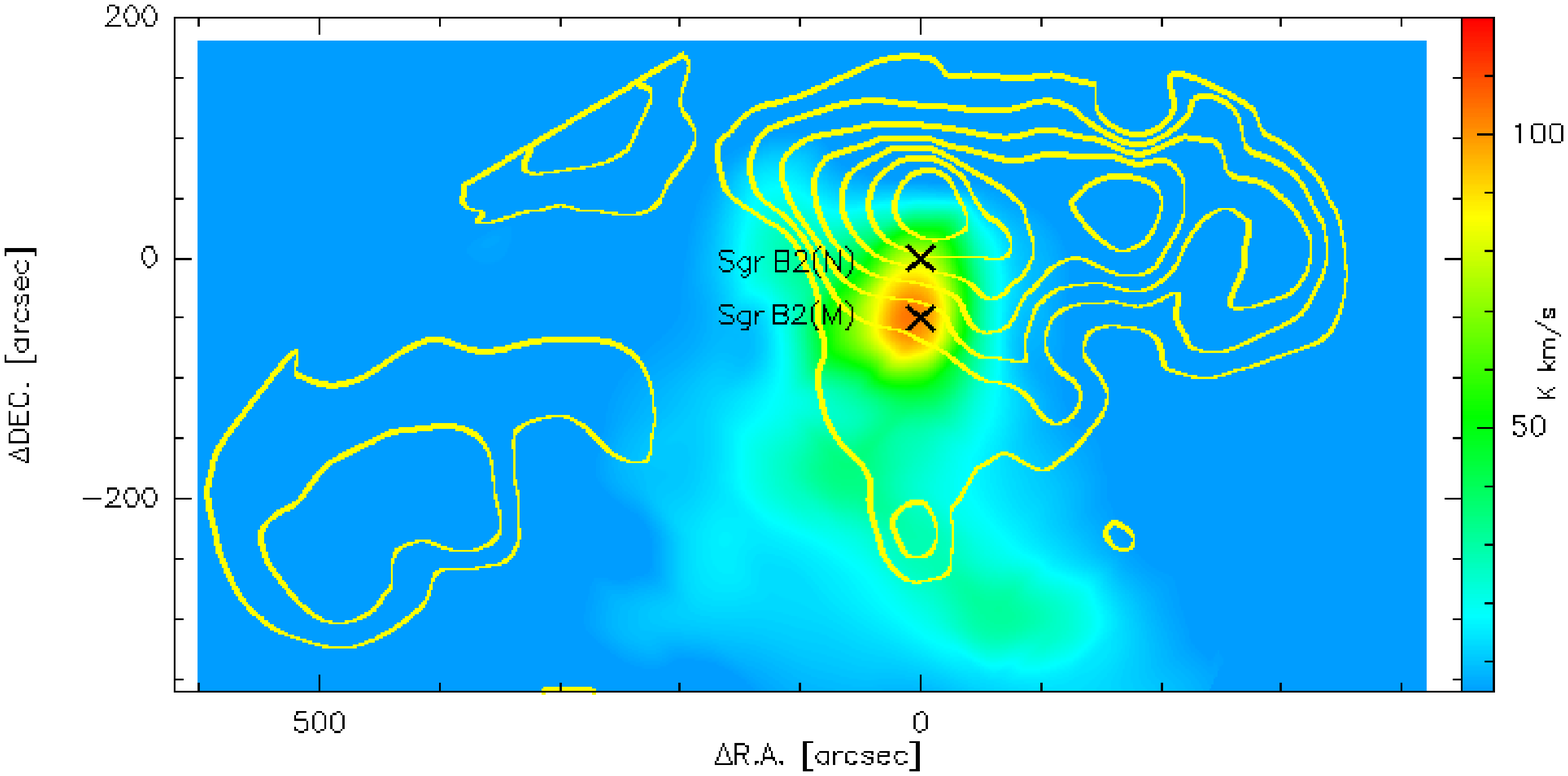} 
 \vspace{-2mm}
\caption{Integrated intensity map of CH$_2$OHCHO $1_{1,0}-1_{0,1}$ (upper panel), HOCH$_2$CH$_2$OH $2_{0,2} (v=0)-1_{0,1} (v=1)$ (middle panel), and HC$_5$N 5-4 (lower panel) emission observed toward Sgr B2 overlaid on H78$\alpha$ emission in color scale. The contours indicate 30\% - 90\%  (steps of 10\%) of the peak integrated intensity, which is 1.08$\pm$0.10 K km s$^{-1}$, 1.06$\pm$0.09 K km s$^{-1}$ and 9.1$\pm$0.2 K km s$^{-1}$ for CH$_2$OHCHO, HOCH$_2$CH$_2$OH and HC$_5$N, respectively. Two crosses are used to denote positions of Sgr B2(N) and Sgr B2(M). }
\end{center}
\end{figure}

\begin{figure}
\begin{center}
\includegraphics[width=0.92\textwidth]{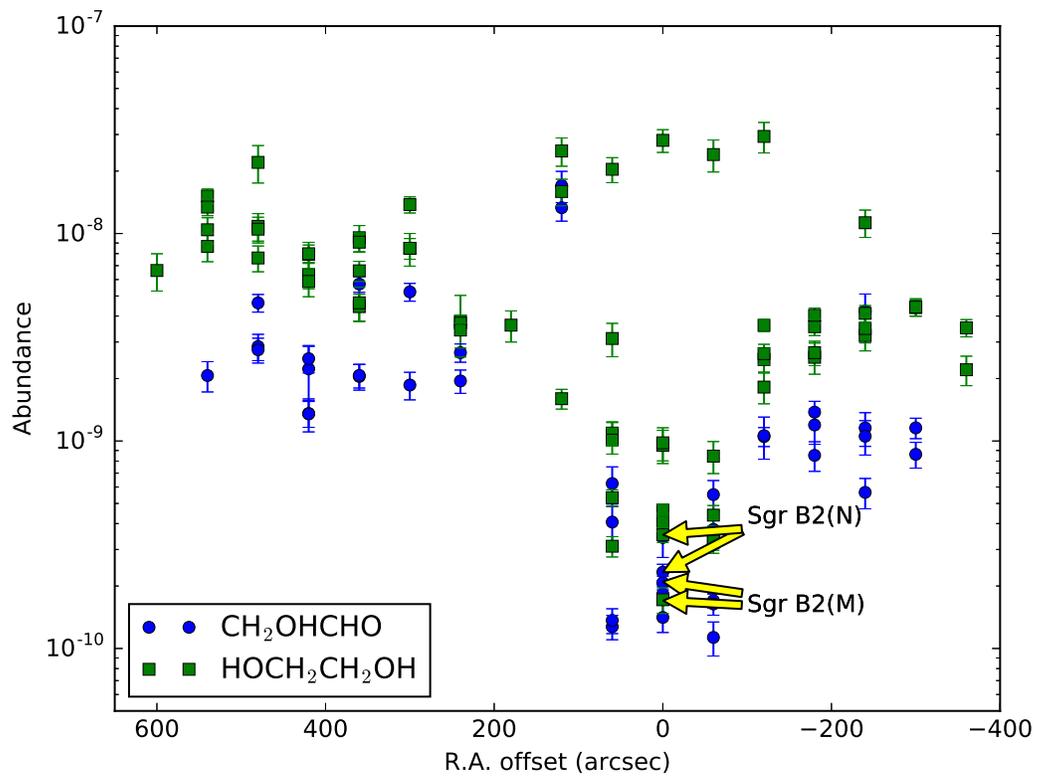} 
\vspace{-2mm}
\caption{Variation of CH$_2$OHCHO (circles) and HOCH$_2$CH$_2$OH (squares)  abundance with R.A. offset away from Sgr B2(N). Results of Sgr B2(N) and Sgr B2(M) were labeled in the figure. }
\end{center}
\end{figure}

\begin{figure}
\begin{center}
\includegraphics[width=0.92\textwidth]{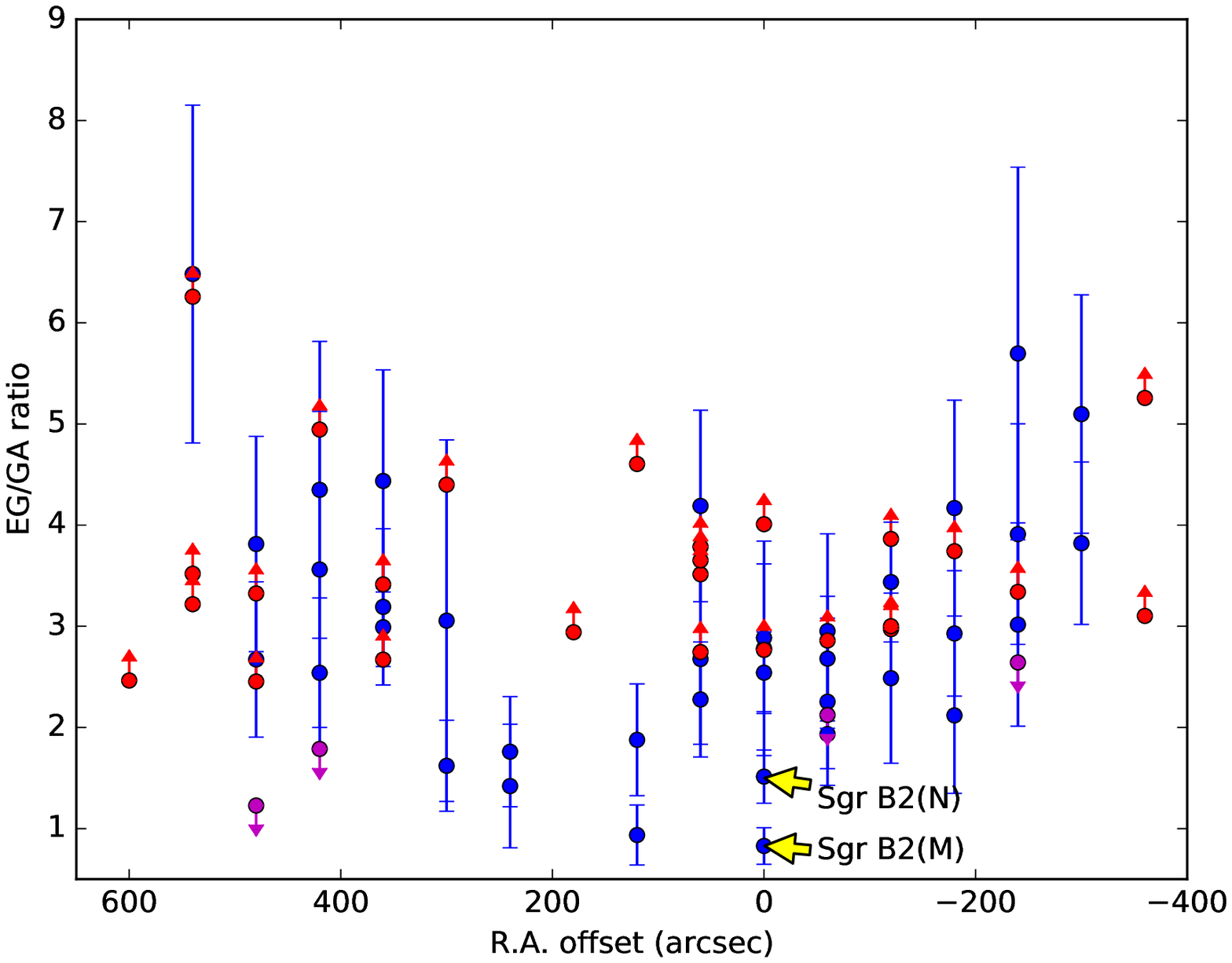} 
\vspace{-2mm}
\caption{Variation of EG/GA abundance ratio with R.A. offset away from Sgr B2(N). Lower limits were given for positions in which only HOCH$_2$CH$_2$OH emission was detected (red dots). Upper limits were given for positions in which only CH$_2$OHCHO emission was detected (magenta dots). Blue dots stand for positions with both of them detected. Results of Sgr B2(N) and Sgr B2(M) were labeled in the figure.}
\end{center}
\end{figure}


\begin{thebibliography}{999} 

\bibitem[An et al. (2017)]{an2017}
An, D., Sellgren, K., Boogert, A.C.A., Ram\' irez, S.V. \& Pyo, T.-S. 2017, ApJ, 843L, 36

\bibitem[bally et al. (2010)]{bally2010}
Bally, J., Aguirre, J., Battersby, C. et al. 2010, ApJ, 721, 137

\bibitem[B et al. (2015)]{balucani2015}
Balucani, N., Ceccarelli, C., Vianney, T. 2015, ApJ, 449, L16 

\bibitem[B et al. (2017)]{barnes2017}
Barnes, A.T., Longmore, S.N., Battersby, C. et al. 2017, MNRAS, 469, 2263

\bibitem[Belloche et al. (2013)]{belloche2013}
Belloche, A., M$\ddot{u}$ller, H.S.P., Menten, K.M., Schilke, P. \& Comito, C. 2013, A\&A, 559, 47

 
\bibitem[Beltran et al. (2009)]{beltran2009}
Beltr$\acute{a}$n, M. T., Codella, C., Viti, S., Neri, R., Cesaroni, R. 2009, ApJ, 690, L93 


\bibitem[B et al. (2006)]{bennett2007}
Bennett, C. J. \& Kaiser, R.I., 2007, ApJ, 661, 899 

  
\bibitem[Biver et al. (2015)]{biver2015}
Biver, N. et~al. 2015, Science Advances, 1, e1500863

 \bibitem[Brouillet et al. (2015)]{brouillet2015}
Brouillet, N. et~al. 2015, A\&A, 576, 129

  
 \bibitem[Butscher et al. (2015)]{butscher2015}
Butscher, T. et~al. 2015, MNRAS, 453, 1587  
  
  \bibitem[Butscher et al. (2016)]{butscher2016}
Butscher, T., Duvernay, F., Danger, G., Chiavassa, T., 2016, A\&A, 593, 60    

 \bibitem[C et al. (2009)]{chapman2009}
Chapman, J.F., Millar, T.J., Wardle, M., Burton, M. G., Walsh, A.J. 2009, MNRAS, 394, 221 


  \bibitem[Chuang et al. (2016)]{chuang2016}
Chuang, K.-J., Fedoseev, G., Ioppolo, S., van Dishoeck, E. F., Linnartz, H. 2016, MNRAS, 455, 1702   

  \bibitem[Chuang et al. (2017)]{chuang2017}
Chuang, K.-J., Fedoseev, G., Qasim, D. et al. 2017, MNRAS, 467, 2552 
    
 \bibitem[Coutens et al. (2015)]{coutens2015}
Coutens, A., Persson, M. V., J$\phi$rgensen, J. K., Wampfler, S. F., Lykke, J. M. 2015, A\&A, 576, 5     

 \bibitem[Coutens et al. (2016)]{coutens2016}
Coutens, A., Rawlings, J.M.C., Viti, S., Williams, D. A. 2017, MNRAS, 467, 737 


  \bibitem[E et al. (2016)]{enrique2016}
Enrique-Romero, J., Rimola, A., Ceccarelli, C., Balucani, N. 2016, MNRAS Letter, 459, 6   
  

 \bibitem[Fedoseev et al. (2015)]{fedoseev2015}
Fedoseev, G., Cuppen, H. M., Ioppolo, S., Lamberts, T., Linnartz, H. 2017, MNRAS, 448, 1288  

 
 \bibitem[Fuente et al. (2014)]{fuente2014}
Fuente, A. et~al. 2014, A\&A, 568, 65   
   

 \bibitem[Garrod et al. (2008)]{garrod2008}
Garrod, R. T., Weaver, S. L. W., Herbst, E. 2008, ApJ, 682, 283 
 
  \bibitem[Geppert et al. (2006)]{geppert2006}
Geppert, W. D. et~al. 2006, Faraday Discussion, 133, 177

 \bibitem[G et al. (2016)]{ginsburg2016}
Ginsburg, A. et~al. 2016, A\&A, 586, 50 

  
 \bibitem[G et al. (2015)]{gong2015}
Gong, Y. et~al. 2015, A\&A, 581, 48  

\bibitem[h et al. (2006)]{halfen2006}
Halfen, D.T., Apponi, A.J., Woolf, N., Plot, R.,   Ziurys, L.M. 2006, A\&A, 639, 237 


\bibitem[Herbst (2009)]{herbst2009}
Herbst, E., van Dishoeck, E. F. 2009,  ARA\&A, 47, 427
  

  
\bibitem[Hollis et al. (2000)]{hollis2000}
Hollis, J. M., Lovas, F. J., Jewell, P. R. 2000, ApJ, 540, L107

 
  \bibitem[Hollis et al. (2001)]{hollis2001}
Hollis, J. M., Vogel, S. N., Snyder, L. E., Jewell, P. R., Lovas, F. J. 2001, ApJ, 554, L81

  \bibitem[Hollis et al. (2002)]{hollis2002}
Hollis, J. M., Lovas, F. J., Jewell, P. R. 2002, ApJ, 571, L59     

  \bibitem[Hollis et al. (2004a)]{hollis2004a}
Hollis, J. M., Jewell, P. R., Lovas, F. J., Remijan, A. 2004, ApJ, 613, L45

  
  \bibitem[J et al. (2016)]{jimenez2016}
Jim\'enez-Serra, I. et~al. 2016, ApJ, 830, L6 

 \bibitem[Jorgensen et al. (2012)]{jorgensen2012}
J${\o}$rgensen, J. K., Favre, C., Bisschop, S.E. et~al. 2012, ApJ, 757, L4  

 
 \bibitem[Jorgensen et al. (2016)]{jorgensen2016}
J${\o}$rgensen, J. K. et~al. 2016, A\&A, 595, 117  

  
\bibitem[K et al. (2016)]{kauffmann2016}
Kauffmann, J. et~al. A\&A, arXiv: 1610.03499
 
       
  \bibitem[Kuan et al. (1996)]{kuan1996}
Kuan, Y.-J., Mehringer, D. M., Snyder, L. E., 1996, ApJ, 459, L619    
 
  \bibitem[Lefloch et al. (2017)]{lefloch2017}
 Lefloch, B. et~al. 2017, MNRAS Letter, 469, 73

  
 \bibitem[Li et al. (2016)]{li2016}
Li, J., Shen, Z.-Q., Wang, J.Z. et~al. 2016, ApJ, 824, 136    

 \bibitem[Lu et al. (2015)]{li2015}
Lu, X., Zhang, Q.Z., Kauffman, J. et al. 2015, ApJ, 814, 18
  
 \bibitem[Lykke et al. (2015)]{lykke2015}
Lykke, J. M., Favre, C., Bergin, E. A., J$\o$rgesen, J. K. 2015, A\&A, 582, 64  
  
 
 
\bibitem[McGuire et al. (2016)]{mcguire2016}
McGuire, B. A. et~al. 2016, Science, 352, 1449     
  
  
\bibitem[Menten (2011)]{menten2011}
Menten, K. M. 2011, ISTP, To apper in the Proceedings of the 4th Cologne-Bonn-Zermatt-Symposium "The Dense Interstellar Medium in Galaxies" eds. S. Pfalzner, C. Kramer, C. Straubmeier, \& Heithausen (Springer: Berlin), 241, 27 



\bibitem[Muller et al. (2005)]{Muller05}
M$\ddot{u}$ller, H. S. P., Schloder, F., Stutzki, J., \& Winnewisser, G. 2005, JMoSt, 742, 215


\bibitem[Oberg et al. (2009)]{oberg2009}
$\ddot{O}$berg, K. I., Garrod, R. T., van Dishoeck, E. F. , Linnartz, H. 2009, A\&A, 504, 891   
  
  
\bibitem[O et al. (2011)]{oberg2011}
$\ddot{O}$berg, K. I. et~al. 2011, IAUS, 280, 65    
  
\bibitem[Palau et al. (2017)]{palau2017}
Palau, A. et~al. 2017, MNRAS, 467, 2723

  
 \bibitem[P et al. (1998)]{pickett1998}
Pickett, R.L. et~al. 1998, J. Quant. Spectrosc. \& Rad. Transfer, 60, 883  
  
\bibitem[P et al. (2005)]{polehampton2005}
Polehampton, E.T., Baluteau, J.-P., Swinyard, B.M. 2005, A\&A, 437, 957


\bibitem[Reid et al. (2004)]{reid2014}
Reid, M.J. et~al. 2014, ApJ, 783, 130
  
  
 \bibitem[Requena-Torres et al. (2008)]{requena2008}
Requena-Torres, M. A., Martin-Pintado, J., Martin, S., Morris, M. R. 2008, ApJ, 672, 352    


 \bibitem[Rivilla et al. (2017)]{rivilla2017}
Rivilla, V.M., Beltr$\acute{a}$n, M. T., Cesaroni, R. et~al. 2017, A\&A, 598, 59

  
\bibitem[sharma et al. (2016)]{sharma2016}
Sharma, M. K., Sharma, A. K., Sharma, M., Chandra, S. 2016, New Astronomy, 45, 45
 
\bibitem[T et al. (2015)]{taquet2015}
 Taquet, V., L$\acute{o}$penz-Sepulcre, A., Ceccarelli, C. et~al. 2015, ApJ, 804, 81   

\bibitem[T et al. (2017)]{taquet2017}
Taquet, V. et~al. A\&A, arXiv:1706.01368   

\bibitem[T et al. (2015)]{tsuboi2015}
Tsuboi, M., Miyazaki, A., Uhehara, K. 2015, PASJ, 67, 109

\bibitem[V et al. (2013)]{vasyunin2013}
Vasyunin, A. I. \& Herbst, E. 2013, ApJ, 769, 34  
  
  
 \bibitem[Woods et al. (2012)]{woods2012}
Woods, P.M. et~al., 2012, ApJ, 750, 19       
 
  
 \bibitem[Woods et al. (2013)]{woods2013}
Woods, P.M. et~al. 2013, ApJ, 777, 90     


\bibitem[Y et al. (2013)]{yusef2013}
Yusef-Zadeh, F., Cotton, W., Viti, S., Wardle, M., Royster, M. 2013, ApJ, 764, L19  
  
\bibitem[Z et al. (2015)]{zhang2015}
Zhang, J. S. et~al. 2015, ApJS, 219, 28  


\end{thebibliography}
\end{document}